\documentclass[a4paper,aps,showpacs,showkeys,twocolumn]{revtex4}
\RequirePackage[english]{babel}
\RequirePackage[latin1]{inputenc}
\RequirePackage[T1]{fontenc}
\RequirePackage{mathrsfs}
\RequirePackage{amsmath}
\RequirePackage{amssymb}
\RequirePackage{amsbsy}
\RequirePackage{color}
\RequirePackage{bm}
\begin{document}
\title{\bf{Torsion Gravity for Dirac Fields}}
\author{Luca Fabbri}
\affiliation{INFN \& Dipartimento di Fisica, Universit{\`a} di Bologna,\\
Via Irnerio 46, 40126 Bologna, ITALY}
\date{\today}
\begin{abstract}
In this article we will take into account the most complete back-ground with torsion and curvature, providing the most exhaustive coupling for the Dirac field: we will discuss the integrability of the interaction of the matter field and the reduction of the matter field equations.
\end{abstract}
\maketitle
\section{Introduction}
In fundamental theoretical physics, torsion-gravity is the theory we obtain when we do not constrain the most general metric-compatible connection to be symmetric in the two lower indices, yielding a geometry that is endowed with torsion as well as the metric, therefore called Riemann-Cartan geometry \cite{C1,C2,C3,C4}; in the Riemann-Cartan geometry the most general metric-compatible connection can be decomposed into the simplest metric-compatible symmetric connection, called Levi-Civita connection and written entirely in terms of the Riemann metric, plus the torsion, called Cartan tensor: in this approach originally followed by Cartan, torsion is just a geometric field whose independence from the metric maintains the most general and the simplest connections correspondingly independent, but it has no other meaning. An additional meaning can be found by writing the Riemann-Cartan geometry in anholonomic bases, in which it becomes more transparent the meaning of torsion as the strength of the potential arising from gauging the translation group much in the same way in which curvature is the strength of the potential arising from gauging the rotation group, as shown by Sciama and Kibble \cite{S,K}: what Sciama and Kibble proved was that torsion is not just something that could be added, but something that must be added, beside curvature, in order to have the possibility to completely describe translations, beside rotations, in a full Poincar\'{e} gauge theory. Sciama and Kibble were also the first to write the dynamics of these fields, by having torsion coupled to the spin much as curvature is coupled to the energy, therefore following in the most general case in which there is also spin, beside energy, the spirit of Einstein gravity. The Sciama-Kibble--Einstein theory of gravity is thus the most complete geometry for the space-time, in which the possibility to couple both spin and energy allows us to have a whole dynamical coupling for spinors, like the Dirac matter field, as reviewed in \cite{h-h-k-n}.

With so much insight, it is an odd circumstance that there be still such a controversy about the role of torsion beside that of curvature, and for an instructive list of fallacies, we refer the reader to reference \cite{Hehl:2007bn}; a selection of topics on torsion in gravity is in \cite{Obukhov:2006gea}. In the present paper, we will consider torsion-gravity as the basis on which to build the most exhaustive coupling to the spin-energy content of spinor fields focusing on the Dirac matter field.

We will see that when both torsion and curvature are taken into account, there will be just enough field equations to have the Dirac field solved in the sense that we will be able to write all physical components of the Dirac field in terms of torsion and curvature, with torsion and curvature undergoing to a system of field equations and constraints in which the Dirac field will no longer appear.
\section{General Dynamics of Matter}

\paragraph{Geometry.} As it has been anticipated just above in the introduction, the most general metric-compatible connection can be decomposed into the simplest metric-compatible symmetric connection, written in terms of the metric, plus the torsion tensor, which is independent on the metric, and we have also stressed that anholonomic bases can be introduced in order to make more manifest the meaning of torsion; but apart from the meaning, there is total equivalence between the formalism of anholonomic (Lorentz) and holonomic (coordinate) bases, and additionally the decomposition of the torsionful connection into the torsionless connection plus torsion makes it clear that all torsionful covariant derivatives and curvatures can be written in terms of the corresponding torsionless covariant derivatives and curvature plus torsional contributions. As a consequence, despite quantities written in the torsionful case in Lorentz formalism have a clearer intuitive significance, nevertheless we may write such quantities either in Lorentz or coordinate formalism and either in the torsionful case or in the torsionless case plus explicit torsional contributions since all of them are mathematically equivalent, and in the following we write all quantities with torsion split away. This will be done for simplicity, and readers interested in the full development can refer to all of the references in the introduction.

On the other hand however in \cite{a-l,m-l} it was discussed how it is reasonable to consider torsion to be completely antisymmetric: here we will assume this is the case and it will be our only assumption, although such constraint amounts to no loss of generality when torsion is coupled to the spin in the case of $\frac{1}{2}$-spin spinors such as the Dirac matter fields, as it is well known; in $(1\!+\!3)$-dimensions, completely antisymmetric rank $3$ tensors are equivalent to the dual of a rank $1$ pseudo-tensor and therefore the torsion tensor can be written in terms of a torsion axial vector $Q^{\alpha\nu\sigma}\varepsilon_{\alpha\nu\sigma\rho}\!=\!W_{\rho}$ with $W_{\rho}$ being the torsion axial vector considered in the following. For future convenience we introduce $\partial_{\alpha}W_{\nu}\!-\!\partial_{\nu}W_{\alpha}\!=\!(\partial W)_{\alpha\nu}$ as the curl of the torsion axial vector; we remind the reader that despite involving partial derivatives nevertheless this expression is a tensor and it is antisymmetric. In the following we will see that it is precisely in terms of the curl of torsion that the dynamics of the torsion field will be assigned.

The metric tensor is given by $g_{\alpha\rho}$ and it will be used to move coordinate indices; tetrads $\xi^{\alpha}_{a}$ are always taken to be ortho-normal $g_{\alpha\rho}\xi^{\alpha}_{a}\xi^{\rho}_{b}\!=\!\eta_{ab}$ and used to pass from coordinate (Greek) indices to Lorentz (Latin) indices: then the Minkowskian matrix $\eta_{ab}$ is used to move the Lorentz indices. The set of Clifford matrices $\boldsymbol{\gamma}^{a}$ verifying
\begin{eqnarray}
&\{\boldsymbol{\gamma}^{a}\!,\!\boldsymbol{\gamma}^{b}\}\!=\! 2\eta^{ab}\boldsymbol{\mathbb{I}}
\end{eqnarray}
are such that from them we define
\begin{eqnarray}
&\frac{1}{4}\!\left[\boldsymbol{\gamma}^{a}
\!,\!\boldsymbol{\gamma}^{b}\right]\!=\!\boldsymbol{\sigma}^{ab}
\end{eqnarray}
as the $\boldsymbol{\sigma}^{ab}$ matrices, which give through
\begin{eqnarray}
&\boldsymbol{\sigma}_{ab}
=-\frac{i}{2}\varepsilon_{abcd}\boldsymbol{\pi}\boldsymbol{\sigma}^{cd}
\end{eqnarray}
the parity-odd matrix $\boldsymbol{\pi}$ (historically, the parity-odd matrix $\boldsymbol{\pi}$ is indicated with the letter gamma and denoted with an index five because it was introduced to study five-dimensional extensions, but because no higher dimension will be considered then such a notation has no longer any meaning and we prefer to adopt a notation in which no index is in display so to avoid confusion): together they verify the anticommutation and commutation relations
\begin{eqnarray}
&\{\boldsymbol{\pi},\boldsymbol{\gamma}_{a}\}=0\\
&\{\boldsymbol{\gamma}_{i},\boldsymbol{\sigma}_{jk}\}
=i\varepsilon_{ijkq}\boldsymbol{\pi}\boldsymbol{\gamma}^{q}
\end{eqnarray}
and
\begin{eqnarray}
&[\boldsymbol{\pi},\boldsymbol{\sigma}_{ab}]=0\\
&[\boldsymbol{\gamma}_{a},\boldsymbol{\sigma}_{bc}]
=\eta_{ab}\boldsymbol{\gamma}_{c}\!-\!\eta_{ac}\boldsymbol{\gamma}_{b}\\
&[\boldsymbol{\sigma}_{ab},\boldsymbol{\sigma}_{cd}]
=\eta_{ad}\boldsymbol{\sigma}_{bc}\!-\!\eta_{ac}\boldsymbol{\sigma}_{bd}
\!+\!\eta_{bc}\boldsymbol{\sigma}_{ad}\!-\!\eta_{bd}\boldsymbol{\sigma}_{ac}
\end{eqnarray}
and also the expressions
\begin{eqnarray}
&\boldsymbol{\gamma}_{a}\boldsymbol{\gamma}_{b}
\!=\!\eta_{ab} \boldsymbol{\mathbb{I}}\!+\!2\boldsymbol{\sigma}_{ab}\\
&\boldsymbol{\gamma}_{i}\boldsymbol{\gamma}_{j}\boldsymbol{\gamma}_{k}
=\boldsymbol{\gamma}_{i}\eta_{jk}-\boldsymbol{\gamma}_{j}\eta_{ik}+\boldsymbol{\gamma}_{k}\eta_{ij}
+i\varepsilon_{ijkq}\boldsymbol{\pi}\boldsymbol{\gamma}^{q}
\end{eqnarray}
valid in general. With $\boldsymbol{\gamma}_{0}$ we define $\overline{\psi}\!=\!\psi^{\dagger}\boldsymbol{\gamma}_{0}$ as the conjugation of spinor fields, and so that we have the possibility to define the bi-linear spinor quantities as
\begin{eqnarray}
&2i\overline{\psi}\boldsymbol{\sigma}^{ab}\psi\!=\!S^{ab}\\
&\overline{\psi}\boldsymbol{\gamma}^{a}\boldsymbol{\pi}\psi\!=\!V^{a}\\
&\overline{\psi}\boldsymbol{\gamma}^{a}\psi\!=\!U^{a}\\
&i\overline{\psi}\boldsymbol{\pi}\psi\!=\!\Theta\\
&\overline{\psi}\psi\!=\!\Phi
\end{eqnarray}
and because of the validity of expression
\begin{eqnarray}
&\psi\overline{\psi}\!\equiv\!\frac{1}{4}\Phi\boldsymbol{\mathbb{I}}
\!+\!\frac{1}{4}U_{a}\boldsymbol{\gamma}^{a}
\!+\!\frac{i}{4}S_{ab}\boldsymbol{\sigma}^{ab}
\!-\!\frac{1}{4}V_{a}\boldsymbol{\gamma}^{a}\boldsymbol{\pi}
\!-\!\frac{i}{4}\Theta \boldsymbol{\pi}\label{F}
\end{eqnarray}
we also have the following relationships
\begin{eqnarray}
&S_{ab}\Phi\!+\!\frac{1}{2}\varepsilon_{abik}S^{ik}\Theta
\!=\!U^{j}V^{k}\varepsilon_{jkab}\label{s}\\
&S_{ab}\Theta\!-\!\frac{1}{2}\varepsilon_{abik}S^{ik}\Phi
\!=\!U_{[a}V_{b]}\label{sdual}
\end{eqnarray}
together with
\begin{eqnarray}
&S_{ik}U^{i}=\Theta V_{k}\label{P1}\\
&-\frac{1}{2}\varepsilon_{abik}S^{ab}U^{i}\!=\!\Phi V_{k}\label{L1}\\
&S_{ik}V^{i}=\Theta U_{k}\label{P2}\\
&-\frac{1}{2}\varepsilon_{abik}S^{ab}V^{i}\!=\!\Phi U_{k}\label{L2}
\end{eqnarray}
and
\begin{eqnarray}
&\frac{1}{2}S_{ab}S^{ab}\!=\!\Phi^{2}\!-\!\Theta^{2}\label{norm2}\\
&U_{a}U^{a}\!=\!-V_{a}V^{a}\!=\!\Theta^{2}\!+\!\Phi^{2}\label{norm1}\\
&\frac{1}{4}S_{ab}S_{ij}\varepsilon^{abij}\!=\!2\Theta\Phi\label{orthogonal2}\\
&V_{a}U^{a}\!=\!0\label{orthogonal1}
\end{eqnarray}
called Fierz re-arrangements and being spinor identities.

Spinors have $8$ real components but being defined in terms of transformation laws with $6$ parameters it is possible to remove some of them reducing spinors to no more than $2$ real components, so if $\overline{\psi}\psi\!=\!i\overline{\psi}\boldsymbol{\pi}\psi\!=\!0$ we have two constraints, meaning that either the spinor has no component left or that we can not perform all of the Lorentz transformations, and anyway the spinor would turn out to be singular; singular spinors are very interesting and in fact they are broadly studied \cite{daRocha:2008we, daSilva:2012wp, Fabbri:2010pk, Vignolo:2011qt, daRocha:2013qhu, Cavalcanti:2014wia, Cavalcanti:2014uta, Fabbri:2011ha, Fabbri:2014wda, Ablamowicz:2014rpa, Fabbri:2012yg, Villalobos:2015xca, Fabbri:2014kea}, but they need particular care for the dynamical features they should have and we are not going to consider them in what follows.

A first comment we have is that although the bi-linear spinor quantities are $16$ linearly independent real tensors, nevertheless they are not independent: as a matter of fact by combining (\ref{s}-\ref{sdual}) one gets the final relationship
\begin{eqnarray}
&S_{ab}\!=\!(\Phi^{2}\!+\!\Theta^{2})^{-1}(U^{j}V^{k}\varepsilon_{jkab}\Phi\!+\!U_{[a}V_{b]}\Theta)
\label{rel}
\end{eqnarray}
showing that the antisymmetric tensor can be obtained so soon as one has the vector and axial-vector as well as the scalar and pseudo-scalar verifying (\ref{norm1}, \ref{orthogonal1}); because at least one between scalar and pseudo-scalar does not vanish then the vector and axial-vector are respectively time-like and space-like and they are always orthogonal.

For general considerations about completely antisymmetric torsion and its consequences for the coupling to the spin of spinors, parity-conservation and continuity in the case of the torsionless limit, we will refer to \cite{Fabbri:2006xq,Fabbri:2008rq,Fabbri:2011kq,Fabbri:2012ag, Fabbri:2013gza,Fabbri:2014dxa}.

For the kinematics of the torsionless quantities we start by specifying that $\Lambda^{\sigma}_{\alpha\nu}$ is the Levi-Civita symmetric connection; with it $\Omega^{a}_{b\pi}\!=\!\xi^{\nu}_{b}\xi^{a}_{\sigma}(\Lambda^{\sigma}_{\nu\pi}\!-\!\xi^{\sigma}_{i}\partial_{\pi}\xi_{\nu}^{i})$ will be the spin connection in general. Then it is possible to define
\begin{eqnarray}
&\boldsymbol{\Omega}_{\mu}
=\frac{1}{2}\Omega^{ab}_{\phantom{ab}\mu}\boldsymbol{\sigma}_{ab}
\!+\!iqA_{\mu}\boldsymbol{\mathbb{I}}\label{spinorialconnection}
\end{eqnarray}
in terms of the spin connection and the gauge potential of charge $q$ as what is called spinorial connection \cite{Fabbri:2015xga}.

As it has been recently proven in \cite{Fabbri:2016msm}, the fact that in general $\Phi^{2}
\!+\Theta^{2}\!\neq\!0$ means that we can always find a frame in which the most general spinor is with no loss of generality reduced to the third-axis rotation eigen-state, and despite there is a total of $4$ such forms nevertheless they are paired by third-axis reflections and in terms of the transform $\psi\!\rightarrow\!\boldsymbol{\pi}\psi$ so only a single form is independent.

This form can be chosen in an arbitrary manner and we choose it to be given in chiral representation as
\begin{eqnarray}
&\!\!\psi\!=\!\exp{(-i\boldsymbol{\pi}\frac{\varphi}{2})}\!\left(\!\begin{tabular}{c}
$\phi$\\
$0$\\
$\phi$\\
$0$
\end{tabular}\!\right)\!=\!\left(\!\begin{tabular}{c}
$e^{i\frac{\varphi}{2}}$\\
$0$\\
$e^{-i\frac{\varphi}{2}}$\\
$0$
\end{tabular}\!\right)\!\phi
\label{spinor}
\end{eqnarray}
showing that left-handed and right-handed semi-spinorial chiral components are the complex conjugate of one another, and giving the bi-linear spinor quantities
\begin{eqnarray}
&V^{3}\!=\!2\phi^{2}\label{a1}\\
&U^{0}\!=\!2\phi^{2}\label{a2}\\
&\Theta\!=\!2\phi^{2}\sin{\varphi}\label{b1}\\
&\Phi\!=\!2\phi^{2}\cos{\varphi}\label{b2}
\end{eqnarray}
where $\phi$ is the module and $\varphi$ is the Takabayashi angle.

The module and the Takabayashi angle are the two degrees of freedom of the spinor as they are the components that cannot be affected by a spinorial transformation, while the redundancy of all other components has been transferred away from the spinor by means of spinorial transformations into the frame; because spinors are fields then the spinorial transformation is also point dependent and the frames are locally defined, and non-inertial frames correspond to non-zero contributions within the spinorial connection: even though the spinor has been left with the essential information alone, nevertheless now we have more complicated spinorial covariant derivatives.

We notice from (\ref{rel}) that the antisymmetric tensor is not necessary given the vector and axial-vector and the scalar and pseudo-scalar; vectors and axial-vectors can always be written in terms of their module times the unitary vector and axial-vector $u^{a}$ and $v^{a}$ identifying their respective directions: from (\ref{a1}, \ref{a2}) and (\ref{b1}, \ref{b2}) we have
\begin{eqnarray}
&V^ {a}\!=\!(\Phi^{2}\!+\!\Theta^{2})^{\frac{1}{2}}v^{a}\label{aux1}\\
&U^{a} \!=\!(\Phi^{2}\!+\!\Theta^{2})^{\frac{1}{2}}u^{a}\label{aux2}
\end{eqnarray}
showing that the vector and axial-vector point to a privileged direction but otherwise they are not necessary given the scalar and pseudo-scalar. We recall that the vector encodes information about the velocity while the axial-vector encodes information about the spin of the matter distribution, so the above reduction actually means that we have boosted in the frame in which the matter distribution is at rest and rotated in such a way that the spin is aligned with the third axis; we also recall that in the above-mentioned reference we proved how a consistently implemented non-relativistic limit requires not only small spatial part of the vector but also small pseudo-scalar, and because the vector is related to the velocity then the pseudo-scalar comes to be related to internal motions of the matter distribution. This interpretation in terms of which the pseudo-scalar is related to intrinsic structures is to be taken together with the fact that the scalar is known to be related to field distributions, and in fact in standard representation the pseudo-scalar describes the interaction between upper and lower parts and the scalar describes the absolute value of upper and lower parts.

\

\paragraph{Dynamics.} For the system of fields in which we are interested, in \cite{Fabbri:2014dxa} we have discussed what is the most general action, and in the following we will consider such an action restricted for simplicity to the least-order derivative terms: in this form it is given in terms of the curl of the Cartan torsion tensor $(\partial W)_{\alpha\nu}$ and the Ricci curvature tensor $R_{\mu\nu}$ with the curl of the Maxwell gauge potential $F_{\alpha\nu}$ and together with the pair of Dirac conjugate spinors $\overline{\psi}$ and $\psi$ according to the expression
\begin{eqnarray}
\nonumber
&\mathscr{L}\!=\!\frac{1}{4}(\partial W)^{2}\!-\!\frac{1}{2}M^{2}W^{2}+\\
\nonumber
&+\frac{1}{k}R_{\mu\nu}g^{\mu\nu}\!+\!\frac{2}{k}\Lambda+\\
\nonumber
&+\frac{1}{4}F^{\alpha\nu}F_{\alpha\nu}-\\
\nonumber
&-i\overline{\psi}\boldsymbol{\gamma}^{\mu}\boldsymbol{\nabla}_{\mu}\psi+\\
&+X\overline{\psi}\boldsymbol{\gamma}^{\mu}\boldsymbol{\pi}\psi W_{\mu}
\!+\!m\overline{\psi}\psi
\label{l}
\end{eqnarray}
where $X$ gives the strength of the interaction between torsion and spin of spinors while $\Lambda$, $M$ and $m$ are the cosmological constant and the masses of torsion and spinor.

When such a least-order derivative action is taken into account by varying it with respect to the metric and torsion fields, the gauge fields and the spinorial fields, one obtains that the gravitational field equations are given according to the following symmetric tensor equations
\begin{eqnarray}
\nonumber
&R^{\rho\sigma}\!-\!\frac{1}{2}Rg^{\rho\sigma}\!-\!\Lambda g^{\rho\sigma}
\!=\!\frac{k}{2}[M^{2}(W^{\rho}W^{\sigma}\!\!-\!\!\frac{1}{2}W^{\alpha}W_{\alpha}g^{\rho\sigma})+\\
\nonumber
&+\frac{1}{4}(\partial W)^{2}g^{\rho\sigma}
\!-\!(\partial W)^{\sigma\alpha}(\partial W)^{\rho}_{\phantom{\rho}\alpha}+\\
\nonumber
&+\frac{1}{4}F^{2}g^{\rho\sigma}\!-\!F^{\rho\alpha}\!F^{\sigma}_{\phantom{\sigma}\alpha}+\\
\nonumber
&+\frac{i}{4}(\overline{\psi}\boldsymbol{\gamma}^{\rho}\boldsymbol{\nabla}^{\sigma}\psi
\!-\!\boldsymbol{\nabla}^{\sigma}\overline{\psi}\boldsymbol{\gamma}^{\rho}\psi+\\
\nonumber
&+\overline{\psi}\boldsymbol{\gamma}^{\sigma}\boldsymbol{\nabla}^{\rho}\psi
\!-\!\boldsymbol{\nabla}^{\rho}\overline{\psi}\boldsymbol{\gamma}^{\sigma}\psi)-\\
&-\frac{1}{2}X(W^{\sigma}V^{\rho}\!+\!W^{\rho}V^{\sigma})]
\end{eqnarray}
whereas the torsional field equations are obtained to be given in terms of the axial-vector equations
\begin{eqnarray}
&\!\nabla_{\alpha}(\partial W)^{\alpha\mu}\!+\!M^{2}W^{\mu}\!=\!XV^{\mu}
\end{eqnarray}
and the gauge field equations are the vector equations
\begin{eqnarray}
&\nabla_{\sigma}F^{\sigma\mu}\!=\!qU^{\mu}
\end{eqnarray}
accounting for all field equations describing how geometry couples to spinors; the spinor field equation is
\begin{eqnarray}
&i\boldsymbol{\gamma}^{\mu}\boldsymbol{\nabla}_{\mu}\psi
\!-\!XW_{\sigma}\boldsymbol{\gamma}^{\sigma}\boldsymbol{\pi}\psi
\!-\!m\psi\!=\!0
\end{eqnarray}
which can be multiplied by $\mathbb{I}, \boldsymbol{\pi}, \boldsymbol{\gamma}_{a}, \boldsymbol{\gamma}_{b}\boldsymbol{\pi}, \boldsymbol{\sigma}_{ab}$ and the conjugate spinor, so that after splitting real and imaginary parts, we obtain the validity of following decompositions
\begin{eqnarray}
&\frac{i}{2}(\overline{\psi}\boldsymbol{\gamma}^{\mu}\boldsymbol{\nabla}_{\mu}\psi
\!-\!\boldsymbol{\nabla}_{\mu}\overline{\psi}\boldsymbol{\gamma}^{\mu}\psi)
\!-\!XW_{\sigma}V^{\sigma}\!-\!m\Phi\!=\!0\\
&\boldsymbol{\nabla}_{\mu}U^{\mu}\!=\!0
\end{eqnarray}
\begin{eqnarray}
&\frac{i}{2}(\overline{\psi}\boldsymbol{\gamma}^{\mu}\boldsymbol{\pi}\boldsymbol{\nabla}_{\mu}\psi
\!-\!\boldsymbol{\nabla}_{\mu}\overline{\psi}\boldsymbol{\gamma}^{\mu}\boldsymbol{\pi}\psi)
\!-\!XW_{\sigma}U^{\sigma}\!=\!0\\
&\boldsymbol{\nabla}_{\mu}V^{\mu}\!-\!2m\Theta\!=\!0
\end{eqnarray}
\begin{eqnarray}
\nonumber
&\frac{i}{2}(\overline{\psi}\boldsymbol{\nabla}^{\alpha}\psi
\!-\!\boldsymbol{\nabla}^{\alpha}\overline{\psi}\psi)
\!-\!\frac{1}{2}\boldsymbol{\nabla}_{\mu}S^{\mu\alpha}-\\
&-\frac{1}{2}XW_{\sigma}S_{\mu\nu}\varepsilon^{\mu\nu\sigma\alpha}\!-\!mU^{\alpha}\!=\!0\\
\nonumber
&\boldsymbol{\nabla}_{\alpha}\Phi
\!-\!2(\overline{\psi}\boldsymbol{\sigma}_{\mu\alpha}\!\boldsymbol{\nabla}^{\mu}\psi
\!-\!\!\boldsymbol{\nabla}^{\mu}\overline{\psi}\boldsymbol{\sigma}_{\mu\alpha}\psi)+\\
&+2X\Theta W_{\alpha}\!=\!0
\end{eqnarray}
\begin{eqnarray}
\nonumber
&\boldsymbol{\nabla}_{\nu}\Theta\!-\!
2i(\overline{\psi}\boldsymbol{\sigma}_{\mu\nu}\boldsymbol{\pi}\boldsymbol{\nabla}^{\mu}\psi\!-\!
\boldsymbol{\nabla}^{\mu}\overline{\psi}\boldsymbol{\sigma}_{\mu\nu}\boldsymbol{\pi}\psi)-\\
&-2X\Phi W_{\nu}\!+\!2mV_{\nu}\!=\!0\\
\nonumber
&(\boldsymbol{\nabla}_{\alpha}\overline{\psi}\boldsymbol{\pi}\psi
\!-\!\overline{\psi}\boldsymbol{\pi}\boldsymbol{\nabla}_{\alpha}\psi)
\!-\!\frac{1}{2}\boldsymbol{\nabla}^{\mu}S^{\rho\sigma}\varepsilon_{\rho\sigma\mu\alpha}+\\
&+2XW^{\mu}S_{\mu\alpha}\!=\!0
\end{eqnarray}
\begin{eqnarray}
\nonumber
&\boldsymbol{\nabla}^{\mu}V^{\rho}\varepsilon_{\mu\rho\alpha\nu}
\!+\!i(\overline{\psi}\boldsymbol{\gamma}_{[\alpha}\!\boldsymbol{\nabla}_{\nu]}\psi
\!-\!\!\boldsymbol{\nabla}_{[\nu}\overline{\psi}\boldsymbol{\gamma}_{\alpha]}\psi)+\\
&+2XW_{[\alpha}V_{\nu]}\!=\!0\\
\nonumber
&\boldsymbol{\nabla}^{[\alpha}U^{\nu]}\!+\!i\varepsilon^{\alpha\nu\mu\rho}
(\overline{\psi}\boldsymbol{\gamma}_{\rho}\boldsymbol{\pi}\!\boldsymbol{\nabla}_{\mu}\psi\!-\!\!
\boldsymbol{\nabla}_{\mu}\overline{\psi}\boldsymbol{\gamma}_{\rho}\boldsymbol{\pi}\psi)-\\
&-2XW_{\sigma}U_{\rho}\varepsilon^{\alpha\nu\sigma\rho}\!-\!2mS^{\alpha\nu}\!=\!0
\end{eqnarray}
which are altogether equivalent to the original spinorial field equation. With them it becomes possible to see that the divergence of the torsion field equations is
\begin{eqnarray}
&M^{2}\nabla_{\mu}W^{\mu}\!=\!2Xm\Theta\label{t}
\end{eqnarray}
while the contraction of the gravitational field equations
\begin{eqnarray}
&-R\!=\!4\Lambda\!+\!\frac{k}{2}\left(-M^{2}W^{2}\!+\!m\Phi\right)\label{c}
\end{eqnarray}
in forms that are indeed considerably simplified.

It is now interesting to reconsider the results that have been found in \cite{Fabbri:2016msm} in the dynamical field equations.

The possibility to always find a frame in which a single spinor field can be written according to (\ref{spinor}) means that we can employ (\ref{b1}, \ref{b2}) and (\ref{t}, \ref{c}) in order to write 
\begin{eqnarray}
&\phi^{2}\sin{\varphi}\!=\!\frac{M^{2}}{4Xm}\nabla_{\mu}W^{\mu}\label{y}\\
&\phi^{2}\cos{\varphi}\!=\!\frac{M^{2}}{2m}W^{2}
\!-\!\frac{1}{km}\left(4\Lambda\!+\!R\right)\label{x}
\end{eqnarray}
or equivalently
\begin{eqnarray}
\nonumber
&\phi^{4}\!=\!\frac{M^{4}}{16m^{2}X^{2}}|\nabla_{\mu}W^{\mu}|^{2}
\!+\!\frac{M^{4}}{4m^{2}}|W^{2}|^{2}-\\
&-\frac{M^{2}}{km^{2}}W^{2}(R\!+\!4\Lambda)\!+\!\frac{1}{k^{2}m^{2}}(R\!+\!4\Lambda)^{2}\label{m}\\
&\tan{\varphi}\!=\!\frac{kM^{2}}{2X}\frac{\nabla W}{kM^{2}W^{2}-2(R+4\Lambda)}\label{a}
\end{eqnarray}
introduced above as the module and the Takabayashi angle, they are shown to be given in terms of the torsion and curvature tensors, and since these are scalar and pseudo-scalar equations then they are valid not only in this frame but in general, and with the very same structural form.

So the form (\ref{spinor}) with equations (\ref{m}, \ref{a}) tells that the spinor field is given in terms of the module and the Takabayashi angle themselves given in terms of the torsion and curvature tensors, which means that when these expressions are re-substituted back into the original system of field equations then all instances of the Dirac field will be replaced by the module and the Takabayashi angle, so ultimately by the torsion and curvature, leaving a system of field equations in which there will only be the torsion and curvature, alongside to the eventual gauge field.

As a consequence of the fact that the Dirac field does not appear any longer in any of the field equations we may claim that the Dirac field has been integrated away and we remark that such an integration is permitted in general by the existence of torsion and curvature.
\section{Rearrangement of Components}

\paragraph{Equations.} In the previous section we have shown how it is always possible to have the spinor field integrated in terms of the torsion and curvature tensors.

When this is done, the entire system of field equations for metric and torsion, gauge fields and spinor fields reduces down to a system of field equations and constraints involving the metric and torsion and gauge fields, that is the spinor field has been integrated in the sense that there is no longer any presence of it in the field equations.

A first thing we would like to notice is that given the identities (\ref{F}), it is possible to show that we have
\begin{eqnarray}
&-V_{a}\boldsymbol{\gamma}^{a}\boldsymbol{\pi}\psi\!=\!U_{a}\boldsymbol{\gamma}^{a}\psi
\!=\!(\Phi\mathbb{I}\!+\!i\Theta\boldsymbol{\pi})\psi\label{FdA1}\\
&-U_{a}\boldsymbol{\gamma}^{a}\boldsymbol{\pi}\psi\!=\!V_{a}\boldsymbol{\gamma}^{a}\psi
\!=\!(\Phi\boldsymbol{\pi}\!+\!i\Theta\mathbb{I})\psi\label{FdA2}
\end{eqnarray}
as well as
\begin{eqnarray}
&iS^{ka}\boldsymbol{\gamma}_{k}\psi
\!=\!U^{a}\psi\!-\!V^{a}\boldsymbol{\pi}\psi\!-\!\Phi\boldsymbol{\gamma}^{a}\psi\label{FdB1}\\
&-\frac{1}{2}\varepsilon^{ijka}S_{ij}\boldsymbol{\gamma}_{k}\psi
\!=\!V^{a}\psi\!-\!U^{a}\boldsymbol{\pi}\psi\!-\!i\Theta\boldsymbol{\gamma}^{a}\psi\label{FdB2}
\end{eqnarray}
which are valid as general spinorial identities.

The spinorial covariant derivative with respect to the spinorial connection (\ref{spinorialconnection}) can be written in the frame in which we have the form (\ref{spinor}) giving the final expression
\begin{eqnarray}
&\!\!\!\!\!\!\!\!i\boldsymbol{\nabla}_{\mu}\psi
\!=\![(i\nabla_{\mu}\ln{\phi}\!-\!qA_{\mu})\mathbb{I}
\!+\!\frac{1}{2}\nabla_{\mu}\varphi\boldsymbol{\pi}
\!+\!\frac{i}{2}\Omega^{ab}_{\phantom{ab}\mu}\boldsymbol{\sigma}_{ab}]\psi\label{scd}
\end{eqnarray}
which is valid in that specific frame but which can next be used to have all instances of the spinorial covariant derivatives substituted in terms of algebraic expressions of the spinor in all the field equations of the above system.

Before proceeding further, we remark that because of the relationship (\ref{rel}) we know that the projections onto the antisymmetric spinorial matrix $\boldsymbol{\sigma}_{ab}$ can be written in terms of the projections onto the $\mathbb{I}, \boldsymbol{\pi}, \boldsymbol{\gamma}_{a}, \boldsymbol{\gamma}_{b}\boldsymbol{\pi}$ matrices.

Therefore we may have the form (\ref{spinor}) plugged into the spinor field equation and the resulting spinor field equation projected onto the above $\mathbb{I}, \boldsymbol{\pi}, \boldsymbol{\gamma}_{a}, \boldsymbol{\gamma}_{b}\boldsymbol{\pi}$ matrices or we may have the form (\ref{spinor}) plugged directly into the first eight of the above spinor field equation's decompositions.

Either way we obtain the validity of the expressions
\begin{eqnarray}
&\!\!\!\!\!\!\!\!(\frac{1}{2}\nabla_{\mu}\varphi
\!+\!\frac{1}{4}\Omega^{\alpha\nu\rho}\varepsilon_{\alpha\nu\rho\mu}
\!-\!XW_{\mu})V^{\mu}\!-\!qA_{\mu}U^{\mu}\!-\!m\Phi\!=\!0\\
&(\nabla_{\mu}\ln{\phi}\!-\!\frac{1}{2}\Omega_{\mu a}^{\phantom{\mu a}a})U^{\mu}\!=\!0
\end{eqnarray}
\begin{eqnarray}
&\!\!\!\!(\frac{1}{2}\nabla_{\mu}\varphi
\!+\!\frac{1}{4}\Omega^{\alpha\nu\rho}\varepsilon_{\alpha\nu\rho\mu}
\!-\!XW_{\mu})U^{\mu}\!-\!qA_{\mu}V^{\mu}\!=\!0\\
&(\nabla_{\mu}\ln{\phi}\!-\!\frac{1}{2}\Omega_{\mu a}^{\phantom{\mu a}a})V^{\mu}
\!-\!m\Theta\!=\!0
\end{eqnarray}
\begin{eqnarray}
\nonumber
&\!\!\frac{1}{2}(\frac{1}{2}\nabla_{\mu}\varphi
\!+\!\frac{1}{4}\Omega^{\alpha\nu\rho}\varepsilon_{\alpha\nu\rho\mu}
\!-\!XW_{\mu})S_{\pi\kappa}\varepsilon^{\pi\kappa\mu\sigma}-\\
&-(\nabla_{\mu}\ln{\phi}\!-\!\frac{1}{2}\Omega_{\mu a}^{\phantom{\mu a}a})S^{\mu\sigma}
\!-\!qA^{\sigma}\Phi\!-\!mU^{\sigma}\!=\!0\\
\nonumber
&(\frac{1}{2}\nabla_{\sigma}\varphi
\!+\!\frac{1}{4}\Omega^{\alpha\nu\rho}\varepsilon_{\alpha\nu\rho\sigma}
\!-\!XW_{\sigma})\Theta-\\
&-(\nabla_{\sigma}\ln{\phi}\!-\!\frac{1}{2}\Omega_{\sigma a}^{\phantom{\sigma a}a})\Phi
\!+\!qA^{\mu}S_{\mu\sigma}\!=\!0\label{int1}
\end{eqnarray}
\begin{eqnarray}
\nonumber
&\!(\frac{1}{2}\nabla_{\sigma}\varphi
\!+\!\frac{1}{4}\Omega^{\alpha\nu\rho}\varepsilon_{\alpha\nu\rho\sigma}
\!-\!XW_{\sigma})\Phi+\\
&+(\nabla_{\sigma}\ln{\phi}\!-\!\frac{1}{2}\Omega_{\sigma a}^{\phantom{\sigma a}a})\Theta
\!-\!\frac{1}{2}qA^{\mu}S^{\pi\kappa}\varepsilon_{\pi\kappa\mu\sigma}
\!+\!mV_{\sigma}\!=\!0\label{int2}\\
\nonumber
&(\frac{1}{2}\nabla_{\mu}\varphi
\!+\!\frac{1}{4}\Omega^{\alpha\nu\rho}\varepsilon_{\alpha\nu\rho\mu}
\!-\!XW_{\mu})S^{\mu\sigma}+\\
&+\frac{1}{2}(\nabla_{\mu}\ln{\phi}\!-\!\frac{1}{2}\Omega_{\mu a}^{\phantom{\mu a}a})
S_{\pi\kappa}\varepsilon^{\pi\kappa\mu\sigma}\!+\!qA^{\sigma}\Theta\!=\!0
\end{eqnarray}
so that from (\ref{int1}, \ref{int2}) and again (\ref{s}-\ref{sdual}) we obtain thus
\begin{eqnarray}
&\!\!\!\!(\nabla_{\mu}\ln{\phi}\!-\!\frac{1}{2}\Omega_{\mu a}^{\phantom{\mu a}a})U^{2}
\!+\!qA^{\rho}U^{\nu}V^{\alpha}\varepsilon_{\mu\rho\nu\alpha}
\!+\!V_{\mu}m\Theta\!=\!0\\
\nonumber
&\!\!(\frac{1}{2}\nabla_{\mu}\varphi
\!-\!\!\frac{1}{4}\varepsilon_{\mu\alpha\nu\rho}\Omega^{\alpha\nu\rho}\!\!-\!\!XW_{\mu})U^{2}+\\
&+qA^{\rho}U_{[\rho}V_{\mu]}\!\!+\!V_{\mu}m\Phi\!=\!0
\end{eqnarray}
which with (\ref{FdA1}, \ref{FdA2}) and (\ref{FdB1}, \ref{FdB2}) are seen to be equivalent to the initial spinor field equation; then the current vector is algebraic and also the spin axial-vector is algebraic but the energy tensor contains spinorial covariant derivatives that can be evaluated with (\ref{scd}), and we obtain
\begin{eqnarray}
&\nabla_{\sigma}F^{\sigma\mu}\!=\!qU^{\mu}
\end{eqnarray}
with
\begin{eqnarray}
&\!\nabla_{\alpha}(\partial W)^{\alpha\mu}\!+\!M^{2}W^{\mu}\!=\!XV^{\mu}
\end{eqnarray}
and
\begin{eqnarray}
\nonumber
&R^{\rho\sigma}\!+\!\Lambda g^{\rho\sigma}\!=\frac{k}{2}[\frac{1}{4}(\partial W)^{2}g^{\rho\sigma}
\!-\!(\partial W)^{\sigma\alpha}(\partial W)^{\rho}_{\phantom{\rho}\alpha}+\\
\nonumber
&+M^{2}W^{\rho}W^{\sigma}\!+\!\frac{1}{4}F^{2}g^{\rho\sigma}
\!-\!F^{\rho\alpha}\!F^{\sigma}_{\phantom{\sigma}\alpha}+\\
\nonumber
&+\frac{1}{8}(\Omega_{ab}^{\phantom{ab}\rho}\varepsilon^{\sigma abk}V_{k}
\!+\!\Omega_{ab}^{\phantom{ab}\sigma}\varepsilon^{\rho abk}V_{k}-\\
\nonumber
&-\Omega_{ijk}\varepsilon^{ijk\sigma}V^{\rho}
\!-\!\Omega_{ijk}\varepsilon^{ijk\rho}V^{\sigma})+\\
\nonumber
&+\frac{1}{2}q|U^{2}|^{-1}(A^{\rho}U^{\sigma}V^{k}V_{k}\!+\!A^{\sigma}U^{\rho}V^{k}V_{k}-\\
\nonumber
&-A_{k}U^{[k}V^{\sigma]}V^{\rho}\!-\!A_{k}U^{[k}V^{\rho]}V^{\sigma})-\\
&\!-m\Phi(\frac{1}{2}g^{\rho\sigma}\!+\!|U^{2}|^{-1}V^{\rho}V^{\sigma})]
\end{eqnarray}
as to complete the system of all of the field equations.

Because what is fundamental are the directions of the vector and axial-vector and the scalar and pseudo-scalar or equivalently the module and Takabayashi angle, then we may re-write the gravitational field equations in form
\begin{eqnarray}
\nonumber
&R^{\rho\sigma}\!+\!\Lambda g^{\rho\sigma}\!=\frac{k}{2}[\frac{1}{4}(\partial W)^{2}g^{\rho\sigma}
\!-\!(\partial W)^{\sigma\alpha}(\partial W)^{\rho}_{\phantom{\rho}\alpha}+\\
\nonumber
&+M^{2}W^{\rho}W^{\sigma}\!+\!\frac{1}{4}F^{2}g^{\rho\sigma}
\!-\!F^{\rho\alpha}\!F^{\sigma}_{\phantom{\sigma}\alpha}+\\
\nonumber
&+\frac{1}{4}\phi^{2}(\Omega_{ab}^{\phantom{ab}\rho}\varepsilon^{\sigma abk}v_{k}
\!+\!\Omega_{ab}^{\phantom{ab}\sigma}\varepsilon^{\rho abk}v_{k}-\\
\nonumber
&-\Omega_{ijk}\varepsilon^{ijk\sigma}v^{\rho}
\!-\!\Omega_{ijk}\varepsilon^{ijk\rho}v^{\sigma})-\\
\nonumber
&-q\phi^{2}(A^{\rho}u^{\sigma}\!+\!A^{\sigma}u^{\rho}+\\
\nonumber
&+A_{k}u^{[k}v^{\sigma]}v^{\rho}\!+\!A_{k}u^{[k}v^{\rho]}v^{\sigma})-\\
&\!-2m\phi^{2}\cos{\varphi}(\frac{1}{2}g^{\rho\sigma}\!+\!v^{\rho}v^{\sigma})]
\end{eqnarray}
the torsion field equations as
\begin{eqnarray}
&\!\nabla_{\alpha}(\partial W)^{\alpha\mu}\!+\!M^{2}W^{\mu}\!=\!2X\phi^{2}v^{\mu}
\end{eqnarray}
and the gauge field equations as
\begin{eqnarray}
&\nabla_{\sigma}F^{\sigma\mu}\!=\!2q\phi^{2}u^{\mu}
\end{eqnarray}
for the interactions, with the spinor field equations as
\begin{eqnarray}
&\!\!\!\!\nabla_{\mu}\ln{\phi}\!-\!\frac{1}{2}\Omega_{\mu a}^{\phantom{\mu a}a}
\!+\!qA^{\rho}u^{\nu}v^{\alpha}\varepsilon_{\mu\rho\nu\alpha}
\!+\!v_{\mu}m\sin{\varphi}\!=\!0\label{F1}\\
&\!\!\!\!\!\!\!\!\!\!\frac{1}{2}\nabla_{\mu}\varphi
\!-\!\!\frac{1}{4}\varepsilon_{\mu\alpha\nu\iota}\Omega^{\alpha\nu\iota}
\!\!+\!qA^{\iota}u_{[\iota}v_{\mu]}\!\!-\!\!XW_{\mu}\!\!+\!v_{\mu}m\cos{\varphi}\!=\!0\label{F2}
\end{eqnarray}
as the form of matter field equations we have eventually.

In this way we have demonstrated, first of all, that all spinorial covariant derivatives could be written in terms of algebraic expressions of the spinor and therefore that all conserved quantities can now be written algebraically, and, secondly, that (\ref{F1}, \ref{F2}) together are equivalent to the expression we obtain when the form (\ref{spinor}) of the spinorial field is substituted into the spinorial field equation.

When combined with the results of the previous section, we see that not only we have the Dirac field integrated away from all field equations but also we have the Dirac field equations re-written as two vector constraints having the structure of real tensors in general.
\section{Internal Structure}
In these first sections we have found that for the spinor having the form (\ref{spinor}) it is possible to have the two scalars integrated away through relationships (\ref{m}, \ref{a}) in terms of torsion and curvature and that with such a form the spinor field equation is equivalent to the pair of vector field equations for the two scalars (\ref{F1}, \ref{F2}), which is to be expected since the two degrees of freedom are determined each with four derivatives and so there have to be exactly two vector field equations as (\ref{F1}, \ref{F2}) actually are.

From now on we focus on the spinor field equations in their equivalent form (\ref{F1}, \ref{F2}) in terms of the vectors
\begin{eqnarray}
&G_{\mu}\!=\!\Omega_{\mu a}^{\phantom{\mu a}a}
\!-\!2qA^{\rho}u^{\nu}v^{\alpha}\varepsilon_{\mu\rho\nu\alpha}\\
&K_{\mu}\!=\!2XW_{\mu}\!+\!\frac{1}{2}\varepsilon_{\mu\alpha\nu\iota}\Omega^{\alpha\nu\iota}
\!-\!2qA^{\iota}u_{[\iota}v_{\mu]}
\end{eqnarray}
so to have them compactly written as
\begin{eqnarray}
&\nabla_{\mu}\ln{\phi^{2}}\!-\!G_{\mu}\!+\!2v_{\mu}m\sin{\varphi}\!=\!0\label{f1}\\
&\nabla_{\mu}\varphi\!-\!K_{\mu}\!+\!2v_{\mu}m\cos{\varphi}\!=\!0\label{f2}
\end{eqnarray}
where $G_{\mu}$ and $K_{\mu}$ are a vector and an axial-vector in terms of torsion, tetrad and gauge fields: they result into second-order field equations for the two scalars given by
\begin{eqnarray}
\nonumber
&\phi^{-2}\nabla^{2}\phi^{2}\!+\!(2m)^{2}+\\
&+2mv^{\mu}(G_{\mu}\sin{\varphi}\!+\!K_{\mu}\cos{\varphi})\!-\!(\nabla_{\mu}G^{\mu}\!+\!G^{2})\!=\!0\\
\nonumber
&\nabla^{2}\varphi\!-\!(2m)^{2}\sin{\varphi}\cos{\varphi}-\\
&-2mv^{\mu}(G_{\mu}\cos{\varphi}\!+\!K_{\mu}\sin{\varphi})\!-\!\nabla_{\mu}K^{\mu}\!=\!0
\end{eqnarray}
the first as a Klein-Gordon equation for the scalar of real mass $2m$ and the second as a sine--Klein-Gordon equation for the pseudo-scalar of imaginary mass $2m$ together with additional disturbing mixing terms; they can be removed by employing the products of (\ref{f1}, \ref{f2}) giving the following
\begin{eqnarray}
&\!\!\left|\nabla \frac{\varphi}{2}\right|^{2}\!-\!m^{2}\!-\!\phi^{-1}\nabla^{2}\phi
\!+\!\frac{1}{2}(\nabla_{\mu}G^{\mu}\!+\!\frac{1}{2}G^{2}\!-\!\frac{1}{2}K^{2})\!=\!0\label{HJ}\\
&\!\nabla_{\mu}(\phi^{2}\nabla^{\mu}\frac{\varphi}{2})
\!-\!\frac{1}{2}(\nabla_{\mu}K^{\mu}\!+\!K_{\mu}G^{\mu})\phi^{2}\label{cont}\!=\!0
\end{eqnarray}
which are much cleaner and they can be recognized to be a Hamilton-Jacobi equation and a continuity equation in which the field $\varphi/2$ can be seen as the action functional and the term given by $\phi^{-1}\nabla^{2}\phi$ is the quantum potential.

The above equations (\ref{f1}, \ref{f2}) determine the behaviour of the matter field: (\ref{f1}) gives in terms of the Takabayashi angle the behaviour of the module; (\ref{f2}) has no module and thus it determines the non-linear behaviour of the Takabayashi angle itself. Only equations (\ref{f2}) display the torsional contribution; both equations have the metric and gauge contributions. These equations account for all derivatives of both fields determining them completely once the boundary conditions are given eventually.

The module $\phi$ squared possesses the usual interpretation of field density for the matter distribution while the Takabayashi angle $\varphi$ being the relative phase between the left-handed and the right-handed semi-spinor fields can be interpreted as containing information about the internal dynamics: in fact, it is precisely the interaction between the two chiral components of the field what gives rise to \textit{Zitterbewegung} effects. Such a phenomenon could be related to the quantum properties of fields.
\section{Thoughts}
An additional remark is about the problem that spinor fields always seemed affected by the presence of negative energies; we believe that this is not necessarily a flaw because no discussion can be definitive about the energy until its coupling to gravity is taken: the only reasonable solution of the problem is to consider the gravitational field equations in their time-time component
\begin{eqnarray}
&R_{00}\!-\!\frac{1}{2}Rg_{00}\!=\!\frac{k}{2}\phi^{2}\Omega_{120}
\end{eqnarray}
where $\phi^{2}\Omega_{120}$ is the energy density, and let the gravitational field equations tell what is the sign of the energy.

All these results are based on the fact that, as we have mentioned above, we have chosen to work with a spinor with spin aligned along the third axis and phase cancelled by a rotation around the same third axis, but of course we could resort to situations in which we do not perform this last third-axis rotation so to leave an overall phase to the spinor field; when this is done and the overall phase is written for plane waves as $e^{-iEt}$ then we have the spin connection given by $\Omega^{12}_{\phantom{12}t}\!=\!2E$ and when the gravitational field is weak enough to have $\xi^{t}_{0}\!\approx\!1$ we see that the energy would turn out to be $2\phi^{2}E$ which is positive defined.

Therefore the energy density should be demonstrated to be positive because that is the property determined by the gravitational field equations, but at least we do know an approximation in which the positivity is ensured.

In rather general terms, because the gravitational field equations have a left-hand side that is non-linear in the spin connection, the gravitational field equations are not symmetric for inversion of the sign of the spin connection itself, and so there are two gravitationally different responses for the same matter field according to whether the energy is positive or negative, and only one can be gravitationally stable; what this means is that even if we could transform a spinor field of positive energy into a spinor field of negative energy, that transformation would not be a symmetry when gravitation is considered.

The problem of negative energy spinors arises only because we retain the validity of a symmetry that is in fact not a symmetry when all field equations are taken.

Additionally, recall that spinors are constituted by two semi-spinors in interaction between each other, with the consequence that if bound states are to exist the internal energy is to be negative: the total energy of the spinor must account for contributions of the internal dynamics, which are negative, beside contributions of the external motion, which are positive, summing up to be undefined.

Positive energy would have to be obtained only in the limit in which the negative energy contribution is lost when neglecting internal dynamics, and in our framework this limit is encoded by the assumption that the evolution of the Takabayashi angle be rendered irrelevant.

Finally, it is interesting to mention what happens in the case of two Dirac fields; we shall recall once more that our analysis exploits the opportunity to perform transformations which transfer degrees of freedom from the spinor to the frame and vice versa: when this is done it is quite easy to understand that the two spinors can be brought to their most simplified form only in two frames that would in general be different, or equivalently since the frame is unique then the reduction can be performed at once on two spinors only if these fields are identical.

And yet another way to see this is by recalling that our results are based on the fact that, as we mentioned above, we chose to work with a spinor whose spin was aligned along the third axis, but of course we could resort to situations in which we do not align the spin along the third axis so to leave the spinor in a form general enough to account for different orientations of the spin; when this is done the consequence is that for the two spinor fields there would be no considerable simplification about the dynamics, and it is solely in the case in which the spinor fields are identical that we can recover the results we have presented here and more information can be obtained.

However, this information tells that when the two identical spinors superpose, the form of the total spinor would merely be twice the form of each single spinor, and because these spinors enter non-linearly in the conserved quantities sourcing the geometric field equations, if each spinor is a solution then their superposition cannot be a solution of the entire system of field equations.

So far as we are concerned, this looks like a dynamical implementation of the exclusion principle.
\section{Conclusion}
In this paper, we have obtained two results regarding the spinor fields having the form (\ref{spinor}): first, the spinor has a total of two physical degrees of freedom represented by the module and the Takabayashi angle which can be integrated by employing (\ref{m}, \ref{a}) in terms of curvature and torsion; second, the spinorial field equation is equivalent to two vectorial equations (\ref{F1}, \ref{F2}) in terms of which the module and the Takabayashi angle have the gradient determined in terms of all other fields. By placing these results together, we have that curvature and torsion have to undergo to the two vectorial equations obtained by substituting (\ref{m}, \ref{a}) into (\ref{F1}, \ref{F2}): they result into two constraints for curvature and torsion. The field equations that are given in form (\ref{F1}, \ref{F2}) or (\ref{f1}, \ref{f2}) are used to infer second-order equations of Klein-Gordon scalar form for the module and sine--Klein-Gordon pseudo-scalar form for the Takabayashi angle, and second-order equations having the structure of the Hamilton-Jacobi equation and the continuity equation suggesting the interpretation for which the module squared represents the field density and the Takabayashi angle represents the action functional.

It has been shown that the module gives rise to the known quantum potential while the Takabayashi angle having opposite sign for the opposite helicities provides information about the interaction between the chiral components; because of this the Takabayashi angle is to be related to \textit{Zitterbewegung} effects. Such a phenomenon could be related to the quantum properties of matter.

These results are important for two reasons: one is that they show how the presence of torsion beside curvature makes it possible to integrate away spinorial degrees of freedom thus going a step forward in the search for exact solutions, and the other is that they make it clear that not only the module but also the Takabayashi angle have to be accounted for a complete description of the dynamical behaviour of \textit{Zitterbewegung} and quantum effects.

In quantum field theory quantum corrections are given by quantizing solutions in form of plane waves but plane waves cannot describe dynamical behaviour related to the presence of spin, while if we considered spin contributions it would be impossible to have the form of plane waves on which quantum corrections are calculated.

But if we considered spin contributions there might be no need to implement any field quantization to obtain the quantum corrections that we observe.
\begin{acknowledgments}
I would like to thank the referee for his comments.
\end{acknowledgments}


\begin{thebibliography}{40}
\bibitem{C1} 
E.Cartan,
\textit{Annales Sci.Ecole Norm.Sup.} \textbf{40}, 325 (1923).
\bibitem{C2} 
E.Cartan,
\textit{Annales Sci.Ecole Norm.Sup.} \textbf{41}, 1 (1924).
\bibitem{C3} 
E.Cartan,
\textit{Annales Sci.Ecole Norm.Sup.} \textbf{42}, 17 (1925).
\bibitem{C4} 
E.Cartan,
\textit{Compt.Rend.Acad.Sci.} \textbf{174}, 593 (1922).
\bibitem{S} 
D.W.Sciama, in \textit{Recent Developments in\\
General Relativity} (Oxford, 1962).
\bibitem{K}
T.W.B.Kibble, 
\textit{J.Math.Phys.} \textbf{2}, 212 (1961).
\bibitem{h-h-k-n}
F.W.Hehl, P.Von Der Heyde, G.D.Kerlick,\\
J.M.Nester, \textit{Rev.Mod.Phys.} \textbf{48}, 393 (1976).
\bibitem{Hehl:2007bn} 
F.W.Hehl, Y.N.Obukhov,\\
\textit{Annales Fond. Broglie.}\textbf{32}, 157 (2007).
\bibitem{Obukhov:2006gea} 
Y.N.Obukhov,
\textit{Int.J.Geom.Meth.Mod.Phys.}\textbf{3}, 95 (2006).
\bibitem{a-l}
J.Audretsch, C.L\"{a}mmerzahl,\\
\textit{Class. Quant. Grav.} \textbf{5}, 1285 (1988).
\bibitem{m-l}
A.Macias, C.L\"{a}mmerzahl,\\
\textit{J. Math. Phys.} \textbf{34}, 4540 (1993).
\bibitem{daRocha:2008we} 
R.da Rocha, J.M.Hoff da Silva,\\
\textit{Adv. Appl. Clifford Algebras}\textbf{20}, 847 (2010).
\bibitem{daSilva:2012wp} 
J.M.Hoff da Silva, R.da Rocha,\\
\textit{Phys. Lett. B}\textbf{718}, 1519 (2013).
\bibitem{Fabbri:2010pk} 
L.Fabbri, S.Vignolo,
\textit{Class.Quant.Grav.}\textbf{28},125002(2011).
\bibitem{Vignolo:2011qt} 
S.Vignolo, L.Fabbri, R.Cianci,\\
\textit{J.Math.Phys.}\textbf{52}, 112502 (2011).
\bibitem{daRocha:2013qhu}
R.da Rocha,L.Fabbri,J.M.Hoff da Silva,R.T.Cavalcanti,\\
J.A.Silva-Neto,\textit{J.Math.Phys.}\textbf{54},102505(2013).
\bibitem{Cavalcanti:2014wia}
R.T.Cavalcanti,
\textit{Int.J.Mod.Phys.D}\textbf{23}, 1444002 (2014).
\bibitem{Cavalcanti:2014uta}
R.T.Cavalcanti, J.M.Hoff da Silva, R.da Rocha,\\
\textit{Eur.Phys.J.Plus}\textbf{129}, 246 (2014).
\bibitem{Fabbri:2011ha} 
L.Fabbri,
\textit{Annales Fond. Broglie}\textbf{38}, 155 (2013).
\bibitem{Fabbri:2014wda} 
L.Fabbri, S.Vignolo, S.Carloni,\\
\textit{Phys.Rev.D}\textbf{90}, 024012 (2014).
\bibitem{Ablamowicz:2014rpa}
R.Ab{\l}amowicz, I.Gon{\c c}alves, R.da Rocha,\\
\textit{J. Math. Phys.}\textbf{55}, 103501 (2014).
\bibitem{Fabbri:2012yg} 
L.Fabbri, S.Vignolo,
\textit{Int.J.Theor.Phys.}\textbf{51},3186(2012).
\bibitem{Villalobos:2015xca}
C.H.Coronado Villalobos, J.M.Hoff da Silva,\\
R.da Rocha, \textit{Eur.Phys.J.C}\textbf{75}, 266 (2015).
\bibitem{Fabbri:2014kea} 
L.Fabbri, P.D.Mannheim,
\textit{Phys.Rev.D}\textbf{90},024042(2014).
\bibitem{Fabbri:2006xq}
L.Fabbri,
\textit{Annales Fond. Broglie.}\textbf{32}, 215 (2007).
\bibitem{Fabbri:2008rq} 
L.Fabbri,
\textit{Annales Fond. Broglie.}\textbf{33}, 365 (2008).
\bibitem{Fabbri:2011kq} 
L.Fabbri,
\textit{Gen.Rel.Grav.}\textbf{45}, 1285 (2013).
\bibitem{Fabbri:2012ag} 
L.Fabbri,
\textit{Int.J.Theor.Phys.}\textbf{53}, 1896 (2014).
\bibitem{Fabbri:2013gza} 
L.Fabbri,
\textit{Gen.Rel.Grav.}\textbf{46}, 1663 (2014).
\bibitem{Fabbri:2014dxa} 
L.Fabbri,
\textit{Int.J.Geom.Meth.Mod.Phys.}\textbf{12},1550099(2015).
\bibitem{Fabbri:2015xga} 
L.Fabbri,
\textit{Gen.Rel.Grav.}\textbf{47}, 119 (2015).
\bibitem{Fabbri:2016msm}
L.Fabbri,
\textit{Int.J.Geom.Meth.Mod.Phys.}\textbf{13},1650078(2016).
\end{thebibliography}
\end{document}